% AMSppt.sty version 2.1.1 is used
\input amstex.tex   \input amsppt.sty

\pageheight{47pc} \vcorrection{-1.1pc} \pagewidth{33pc}
\TagsOnRight
%\baselineskip

\def\change#1\to#2\endchange{{\tt -change-} #1 {\tt -to-} #2 {\tt -end-}}
\def\add#1\endadd{{\tt -add-} #1 {\tt -end-}}

\magnification\magstep1

\PSAMSFonts

\def\RR{{\bold R}}

\def\SS{{\bold S}}
\def\HH{{\bold H}}

\def\Arg{\operatorname{Arg}}
\def\at{\char`@}
\def\Bu{B}

\def\Det{\operatorname{det}}
\def\e{\operatorname{e}}
\def\euc#1#2{\langle#1 \vert #2 \rangle_{{}_E}}
\def\FF{{\Cal F}}

\def\Lt{[L]}%{\widetilde L}}

\def\PP{{\Cal P}}

\def\sign{\operatorname{sign}}
\def\Sn{S}

\def\stress#1{{\it#1\/}}
\def\thetat{[\theta]}%{\widetilde\theta}}

\def\Yt{[Y]}%{\widetilde Y}}

\topmatter

\author Pedro de M. Rios
\endauthor

\address Departamento de Matem\'atica, ICMC, Universidade de
S\~ao Paulo; \hfill\break\indent Cx Postal 668, S\~ao Carlos, SP,
13560-970, Brasil.
\endaddress

\email prios\at icmc.usp.br \endemail

\title On (non)commutative products of functions on the sphere
\endtitle

\abstract We investigate the commutativity of global products of
functions on $\SS^2$ from the point of view of a construction
started in \cite{RT} and named the skewed product. We complete the
construction of the skewed product of functions on $\SS^2$ and
show that it is ${\bold Z_2}$-graded commutative and nontrivial
only as a product of functions with correct parity under the
antipodal mapping. These properties are valid for a more general
class of integral products of functions on the sphere, with
integral kernel of a special WKB-type that is natural from
semiclassical considerations. We argue that our construction
provides a simple geometrical explanation for an old theorem by
Rieffel \cite{Rf} on equivariant strict deformation quantization
of the two-sphere.
\endabstract

\endtopmatter

\refstyle{A}

\document

%\newpage

\head Introduction \endhead

An integral product of functions on a simply connected symplectic
symmetric space $M$ has been defined in \cite{RT}. This product,
which we name {\it skewed product} of functions on $M$, is a
mathematical invention. The motivation for its construction is to
provide a simple geometric framework that allows for generalizing
to some other symplectic manifolds the product of functions on
$\RR^{2n}$ that has been defined by Weyl, von Neumann, Groenewold
and Moyal \cite{Wl}\cite{vN}\cite{Gr}\cite{Ml}, with emphasis on
its integral formulation \cite{vN}\cite{Gr}, which is better
suited to more careful treatments of oscillatory functions
\cite{Rs}.

The inspiration for the construction of the skewed product comes,
on the one hand, from Berry's ``center'' description in
semiclassical mechanics \cite{By} and, on the other hand, from the
prequantization approach developed by Kostant and Souriau
\cite{Ko}\cite{So}. Via symplectic groupoids \cite{WX}, these two
geometric guidelines are brought together.

This name ``skewed product'' has been coined to stress the
distinction of its construction from the better known product
defined via formal deformation quantization, which is named ``star
product'' \cite{B-S}, and the product defined within the context
of quantum theory via symbol mapping homomorphism, which is named
``twisted product'' in \cite{VG} (though these two products are
often confused and named star product in the general literature).

As opposed to twisted products, no Hilbert space structure (and
its quantum theory) is required for the construction of the skewed
product. Also, as opposed to formal deformation quantization,
associativity is not imposed beforehand. Its construction is
purely geometrical, fairly simple, and the product agrees with an
ansatz by Weinstein \cite{Wn} for an integral kernel of a special
WKB-type which is related, on the one hand, to the product of von
Neumann and Groenewold and, on the other hand, to the composition
of central generating functions of canonical relations on
symmetric symplectic spaces \cite{Mv}\cite{RO}.

Accordingly, it should at least be expected that the skewed
product provides valuable information on ``quantum'' products at
the semiclassical limit $(\hbar\to 0^+)$, or as a framework for
studying ``quantizations'' $(0^+ \to \hbar)$ of symmetric
symplectic spaces.

The construction of the skewed product, as presented in \cite{RT},
is well defined and unique when $M$ is a simply connected
symplectic symmetric space of noncompact type, but it is also
possible to be carried out, at least partially, in the compact
case.

In this paper, after reviewing the partial construction in
\cite{RT}, we complete the construction of the skewed product of
functions on the $2$-sphere, showing that the global skewed
product $\star^n$ of functions on $\SS^2$ is ${\bold Z_2}$-graded
commutative, $f\star^n g = (-1)^n g\star^n f$, and is nontrivial
only as a product on the subspace of functions satisfying
$f(-\alpha) = (-1)^nf(\alpha)$, where $n$ is the prequantization
integer and $-\alpha$ is antipodal to $\alpha$ (Theorems 1 and 2).

In fact, the main reason why a well defined skewed product in an
infinite dimensional space of functions on the sphere must satisfy
these properties, also implies that they are satisfied quite
independently of the particular amplitude function for an integral
kernel of WKB-type, just as long as the kernel of WKB-type is
symmetric, $SU(2)$ invariant, and its phase is a half-integral
multiple of a midpoint triangular area (Corollary 3).

In light of an old result by Rieffel \cite{Rf}, we discuss the
sharp contrast between our result and the noncommutative property
of any star product of formal power series in $\hbar$ with
coefficients in the space of smooth functions on the sphere
\cite{MO}\cite{Md}. We argue that our construction provides a
simple geometrical explanation of Rieffel's theorem.

\head The skewed product of functions on the $2$-sphere
\endhead

\subhead (Partial) construction of the skewed product of functions
\endsubhead

The skewed product of functions on a simply connected symmetric
symplectic space $(M, \omega, \nabla)$ has been defined in
\cite{RT}. The defining elements of the construction of the skewed
product are: (i) prequantization of the pair groupoid; (ii)
central polarization; (iii) central polarized sections and central
decomposition; (iv) product of sections and holonomy of the
central decomposition; (v) averaging procedure. We now summarize
the construction:

(i) Denoting by $\rho_{+}$ and $\rho_{-}$ the two canonical
projections $M \times M \to M$, the {\it pair} (symplectic) {\it
groupoid} over $M$ is the manifold $M \times M$ with symplectic
form $\rho_{+}^*\omega - \rho_-^*\omega =:(\omega,-\omega)$.  If
$\pi:Y\to M$ is a prequantization of $M$ with connection $\theta$,
the manifold $Y \times Y$ is a principal torus bundle over $M
\times M$. Taking the quotient of the diagonal action of $\SS^1$,
we obtain a principal $\SS^1$-bundle $\Yt = (Y \times Y) / \SS^1
\to M \times M$. We denote points in $\Yt$ as $[y_1,y_2]$ with
$y_i \in Y$. The induced $\SS^1$-action is $\e^{i\phi}\cdot
[y_1,y_2] = [\e^{i\phi}\cdot y_1,y_2] = [y_1,\e^{-i\phi}\cdot
y_2]$. The 1-form $(\theta,-\theta)$ induces a connection form
$\thetat \equiv [\theta,-\theta]$ on $\Yt$, whose curvature is
$(\omega/\hbar, -\omega/\hbar)$. $\Yt$ is a prequantization of $M
\times M$ which is an $\SS^1$ extension of the pair groupoid, for
the space of unities as the diagonal section $\varepsilon_0:M \to
\Yt$, $\varepsilon_0(m) = [y,y]$ with $y\in Y$ s.t. $\pi(y) = m$,
which is horizontal for the connection $\thetat$, and the groupoid
product reading $[x,y] \odot [y,z] = [x,z]$. We let $\Lt \to M
\times M$ be the associated complex line bundle with connection
and compatible hermitian structure, so that we can identify
$\Yt\subset\Lt$.

(ii) For the complete affine connection $\nabla$ on $TM$, geodesic
inversion is a symplectomorphism, so we define the map: $ F : TM
\to M \times M \ ; \ F(m,v) = (\exp_m(-v), \exp_m(v))$, where
$\exp_m : T_mM \to M$ denotes the geodesic flow at time $t=1$,
starting at $m\in M$, in the direction of $v\in T_mM$. $F$ is a
diffeomorphism in a neighborhood of the zero section of $TM$, so
we define $U \subset TM$ as the maximal connected and open
neighborhood of the zero section on which $F$ \stress{is} a
diffeomorphism, and its image $V = F(U) \subset M\times M$ (if
$\nabla$ has no closed geodesics, then $U=TM$ and $V= M\times M$).
On $TM$ there is a natural foliation $\FF_v$ whose leaves are the
fibres $T_mM$. Restricting $\FF_v$ to $U$ defines a polarization
of $U$, with pullback symplectic form $\Omega=F^*(\omega,
-\omega)$, which we call the {\it vertical} or {\it central
polarization}.

(iii) The section $\varepsilon_0$ of $\Yt$ gets transformed to a
section $\epsilon_0$ of the pullback bundle $(\Bu,\Theta) =
F^*(\Yt,\thetat)$ above $M$ seen as the zero section of $TM$.
Since the fibers $T_mM$ are simply connected and since the
pullback connection $\Theta$ has null curvature on these fibers,
$\epsilon_0$ extends to a global section $\sigma : TM \to \Bu$
which is horizontal on leaves of $\FF_v$.  Denote by $\sigma_0$
the restriction of $\sigma$ to $U$. Pushing it by $F$ we obtain a
section $s_0$ over $V$, which, $\forall(m_1,m_2)\in V$ gives
$s_0(m_1,m_2) = [x,y]$ where $x$ and $y$ are the end points of a
horizontal curve in $Y$ above the geodesic on $M$ between $m_1$
and $m_2$ whose pre-image $F^{-1}(m_1,m_2)$ is in $U$. We call
{\it central polarized sections}, the sections of $\Lt$ above $V$
that are covariantly constant along $\PP=F_*(\FF_v|_U)$. Viewing
$\Yt\subset\Lt$, the section $s_0$ constructed above is
$\PP$-constant and, since $s_0$ is a smooth nowhere vanishing
section, central polarized sections are in bijection to functions
$f$ that are constant on the leaves of $\PP$, i.e., with functions
on $M = V/\PP$. The identification is $s = f \cdot s_o$, or, more
precisely, $s(m_1,m_2) = f(m_{12}) \cdot s_0(m_1,m_2)$, where
$m_{12}$ is the midpoint of the geodesic on $M$, between $m_1$ and
$m_2$, s.t. $F^{-1}(m_1,m_2) = (m_{12},v)\in U$. More generally,
any section $s: V\to\Lt$ can be decomposed as $s = \phi\cdot s_o$,
where $\phi$ is a function on $V$. We call this the {\it central
decomposition}, that separates the phase dependence along the
fibers $T_mM$ which is obtained by parallel transporting the
identity section $\varepsilon_0\cong\epsilon_0$.

(iv) Any $p\in \Lt$ can be written in a unique way as $p = \lambda
[x,y]$ with $\lambda \in [0,\infty)$ and $[x,y] \in \Yt \subset
\Lt$. Now, for $p_i = \lambda_i [x_i,y_i]$ such that $\pi(y_1) =
\pi(x_2)$ we define $ p_1 \odot p_2 = \lambda_1 \lambda_2
[x_1,y_1]\odot[x_2,y_2]$. With this extended quasi-groupoid
structure (quasi because now not every element has an inverse), we
construct a {\it product of two sections} $s^1$ and $s^2$ of $\Lt$
by $ (s^1 \circledcirc s^2)(m_1,m_3) = \int_M s^1(m_1,m_2) \odot
s^2(m_2,m_3) \,dm_2$, where $dm_2$ is the Liouville measure on
$(M,\omega)$. For two $\PP$-constant sections $s^i = f_i\cdot
s_0$, each $f_i$ a function on $M$, we get $ (s^1\circledcirc s^2
)(m_1,m_3) = \int_M f_1(m_{12}) f_2(m_{23}) s_0(m_1,m_2) \odot
s_0(m_2,m_3) \,dm_2$, in which $m_{jk}$ denotes the midpoint of
the geodesic between $m_j$ and $m_k$, with
$F^{-1}(m_j,m_k)=(m_{jk},v_{jk})\in U$. However, $s_0(m_1,m_2)
\odot s_0(m_2,m_3) = \lambda s_0(m_1,m_3)$, where $\lambda\in
\SS^1$ is the {\it holonomy over the geodesic triangle} with
vertices $(m_1,m_2,m_3)$ and all geodesics with pre-images in $U$,
so that ${(f_1\cdot s_0 \circledcirc f_2 \cdot s_0)(m_1,m_3)} =
\phi(m_1,m_3)\cdot s_0(m_1,m_3)$, where $\phi$ is a function on
$V$. In this way, we have associated to two functions $f_1$, $f_2$
on $M$ a new function $\phi$ on $V\subset (M \times M)$.

(v) In order to get a new central polarized section from the
product of two central polarized sections, i.e., in order to
associate to two functions $f_1$ and $f_2$ on $M$ a new function
$f_1 \star f_2$ on $M$, we integrate (average) $\phi$ over the
leaves of $\PP$. In terms of the fibres of $TM$ we get $(f_1 \star
f_2)(m) = \int_{U_{m}} dv\ \phi (F(m,v))$, where $U_m = T_mM \cap
U$. We choose the measure $d_{U_m}(v) \equiv dv$ on $U_m$ s.t.
$d_M(m) \wedge d_{U_m}(v) = d_U(m,v)$, where $d_M(m)$ is the
Liouville measure on $(M,\omega)$ and $d_U(m,v)$ is the Liouville
measure on $(U,\Omega)$. This choice for $dv$ defines our {\it
averaging procedure}. If $\nabla$ has no closed geodesics, $U=TM$
and, in this case, $f_1 \star f_2$ defined above is {\it the
skewed product} of $f_1$ and $f_2$. Otherwise, we call $f_1 \star
f_2$ a {\it partial} skewed product, because we have used only a
(big) neighborhood $U\subset TM$ of the zero section.

\subhead A partial skewed product of functions on the sphere
\endsubhead

Skewed products can be written entirely in terms of midpoint
geometrical data (using that $M$ is prequantized). When
$M=\RR^{2n}$, it coincides with the product of von Neumann and
Groenewold \cite{RT}. Coming to the sphere, let $\alpha, \beta,
\gamma$ be three points on $\SS^2$, also thought as three unit
vectors in $\RR^3$, and let $n=Area(\SS^2)/2\pi\hbar = 2/\hbar \in
{\bold Z^{+}}$ be the prequantization integer. A {\it partial}
skewed product of two functions $f_1$ and $f_2$ on $\SS^2$ is
given by \cite{RT}:
$$
f_1\!\star\! f_2 \ (\gamma) \ = \iint_{W_{\gamma}} f_1(\alpha)
f_2(\beta) \, \e^{in\Sn(\alpha,\beta,\gamma)/2}
A(\alpha,\beta,\gamma) \,
 d\alpha \, d\beta
\ . \tag1
$$

We now identify the various elements in formula \thetag1 above.
First,
$$
\Sn(\alpha,\beta,\gamma) = 2\Arg\Bigl( \eta\sqrt{1-
\Det(\alpha\beta\gamma)^2} + i \Det(\alpha\beta\gamma) \Bigr) \ ,
\tag2
$$
is the symplectic area of a geodesic triangle with
$(\alpha,\beta,\gamma)$ as midpoints and sides whose lengths are
all smaller than or equal to $\pi$. Not all triple of midpoints
satisfy this restriction on the lengths of the sides of the
corresponding midpoint triangle. Given any $\gamma\in\SS^2$, only
those $(\alpha,\beta)$ in $W_{\gamma}\subset\SS^2\times\SS^2$
satisfy this restriction, where
$$
W_\gamma = \{ (\alpha, \beta) \in \SS^2 \times \SS^2 \mid \sign
\euc\alpha\beta = \sign\euc\beta\gamma  =  \sign \euc\gamma\alpha
\, \} \ , \tag3
$$
with $\euc\alpha\beta$ denoting the usual scalar product of
vectors $\alpha,\beta$ in $\RR^3$. Accordingly, in formula
\thetag2 $\eta$ is this sign of the three scalar products (for
triangles with at most one side bigger than $\pi$, then $\eta$ is
the same as the majority of signs among the three scalar
products).

Now, if one (and therefore all) of these three scalar products is
not zero, there is a bijection $G^{-1}$ that takes the three
midpoints $(\alpha,\beta,\gamma)$ in the restricted set above to
the three vertices $(a,b,c)$ of the geodesic triangle with all
sides smaller than $\pi$. Then, $A$ is the jacobian of this
transformation, that is, if $da$, $d\alpha$, etc, is the canonical
measure on $\SS^2$,
$$(G^{-1})^*(dadbdc)=A(\alpha,\beta,\gamma)d\alpha
d\beta d\gamma \ ,
$$
$$
A(\alpha, \beta, \gamma) = 16\, \Bigl\vert
 \euc\alpha\beta \cdot
\euc\beta\gamma \cdot \euc\gamma\alpha \Bigr\vert \cdot \Bigl( 1 -
\Det(\alpha\beta\gamma)^2 \Bigr)^{-5/2} \ . \tag4
$$

We also note that the integral kernel $K=Ae^{in\Sn/2}$ is {\it
symmetric}, in the sense that
$$
K(\alpha,\beta,\gamma)=K(\gamma,\alpha,\beta)=\overline{K}(\gamma,\beta,\alpha)
\ . \tag5
$$
Accordingly, we say that the integral product given by formula
\thetag1 is {\it symmetric}, in this sense, which must not be
confused with commutative. This symmetric property for the skewed
product is general and not particular for the case of the sphere.
The same can be said of the geometrical interpretation for the
phase, the amplitude and the domain of integration of the skewed
product, which are valid in general \cite{RT}.

Finally, we say that the integral product given by formula
\thetag1 defines a {\it restricted} skewed product, because the
integration is carried over a proper subset of $\SS^2\times\SS^2$.
This is also true of {\it the} skewed product of functions on the
noncompact hyperbolic plane $\HH^2$ \cite{RT}, even though that
skewed product is not partial because, in that case, $U=T\HH^2$.

\subhead Extending the construction: antipodal midpoints
\endsubhead

We now notice that the restriction \thetag3 is unnatural. This is
so because the amplitude function \thetag4 does not depend on this
restriction, in sharp contrast with the relation between the
amplitude and the restriction for the case of the hyperbolic plane
\cite{RT}. Thus, we question if the product \thetag1 is unique, if
it can be written equivalently in other domains, or if it can be
nontrivially extended to a larger domain in $\SS^2 \times \SS^2$
in a well defined and unique way. We shall investigate these
questions from the point of view of a natural generalization of
the averaging procedure. To do so, we must deal with the lack of
uniqueness in determining spherical triangles, as determined by
midpoints. This, in turn, is rooted in the non uniqueness of
geodesics connecting any two points on the sphere, not just those
that are antipodal.

If two points on $\SS^2$ stand in antipodal relation, there is an
infinity of geodesics, with an infinity of directions, connecting
these points. Then, the whole equator of these points is the set
of their possible midpoints. On the other hand, if two points $a$
and $b$ in $\SS^2$ do not stand in antipodal relation, there still
exists an infinity of geodesics connecting these points, but they
all have the same direction. Thus, there are only two possible
midpoints for this pair of points: the midpoint $\alpha$ of the
shortest geodesic connecting $a$ to $b$ and its antipodal
$-\alpha$, which is the midpoint of the geodesic that composes
with the shortest one (with reverse orientation) to form a big
circle on the sphere. Borrowing from the terminology in \cite{RO},
we say that these two geodesics are weakly equivalent to each
other.

All other geodesics connecting $a$ and $b$ are strongly equivalent
to one of these, in the sense that their midpoints coincide. In
what follows, it will become clear that strongly equivalent
geodesics can be treated as the same, therefore we shall refer to
these two strongly inequivalent geodesics as {\it the short} and
{\it the long} geodesic connecting $a$ to $b$.

Looking at $V\subset(\SS^2 \times \SS^2)$, the set of all pairs of
points in $\SS^2$ that are not antipodal, we can identify the
pre-image $U\equiv U_0\subset T\SS^2$ as the set of all tangent
vectors $\tau=(m,v)$ whose lengths $|v|$ are smaller than $\pi/2$.
However, we can also identify another pre-image in $T\SS^2$, $U_1
\neq U \equiv U_0$, which is the set of all tangent vectors
$\tau=(m,v)$ whose lengths $|v|$ satisfy $\pi/2 < |v| \leq \pi$.
Although $F(U_1) = V$, the inverse is ill defined because all
vectors of length $\pi$ based at $\alpha$ have as image the pair
$(-\alpha,-\alpha)$. In order to define a bijection, we must
identify all such tangent vectors $\tau_1$ and $\tau_2$ under the
equivalence relation $\tau_1 \sim \tau_2$ if $m_1=m_2$ and
$|v_1|=|v_2|=\pi$. Then, $\widetilde{U_1} = (U_1/\sim)$ stands in
bijection to $V$ and thus to $U\equiv U_0$. In what follows, it
will become clear that $\widetilde{U_1}$ is the only other space
that needs to be considered for generalizing our construction.
Clearly, the vertical polarization is natural to
$\widetilde{U_1}$, but remember that, if $\alpha$ is the base of
the element $\tau\in U_0$, s.t. $F(\tau) = (a,b)\in V$, then
$-\alpha$ is the projection of $\tau'\in\widetilde{U_1}$, s.t.
$F(\tau') = (a,b)\in V$.

\subhead Unique composition of central polarized sections
\endsubhead

The next step of the construction in \cite{RT} to be generalized,
pulling back the prequantized bundle to $\widetilde{U_1}$, depends
on the choice of a trivializing section $\epsilon_1$ over the new
pre-image of the diagonal in $\SS^2 \times \SS^2$ and its
extension to a horizontal section $\sigma_1$ over
$\widetilde{U_1}$ that pushes forward under $F$ to two new
sections $\varepsilon_1:M \to \Yt$ and $s_1 : V \to \Yt$,
respectively.

The natural choice for $\epsilon_1$ and its extension to
$\sigma_1$ is explained in proposition 6.1' of \cite{RO} and is
such that, as $\sigma_0$ is the restriction to $U\equiv U_0
\subset T\SS^2$ of a single section $\sigma : T\SS^2 \to B$,
$\sigma_1$ is the restriction of this same section $\sigma$ to
$\widetilde{U_1}$, in such a way that the pushed forward section
$s_1$ satisfies $s_1(a,b) = [x',y']$, where $x'$ and $y'$ are the
endpoints of a horizontal curve above the long geodesic between
$a$ and $b$.

The fact that $\sigma_1$ is well defined on $\widetilde{U_1}$, not
just on $U_1$, is a consequence of the prequantization condition
so that, given the positive integer $ n=Area(\SS^2)/2\pi\hbar =
2/\hbar \in {\bold Z^{+}}$, it is clear that $\varepsilon_1(a)
\cong s_1(a,a) = (-1)^n\cdot s_0(a,a) \cong (-1)^n\cdot
\varepsilon_0(a)$, since $(-1)^n$ is the holonomy along any great
circle. It follows that $\epsilon_1(-a,[\pi])=
(-1)^n\cdot\epsilon_0(a,0)$ and this relation extends naturally so
that, if $\tau\in U_0$ and $\tau'\in\widetilde{U_1}$ are s.t.
$F(\tau) = F(\tau') = (a,b)\in V$, then
$$
s_1(a,b)=(-1)^n\cdot s_0(a,b) \ .
$$

Now, $\PP = F_*(\FF_v|_{U_0})=F_*(\FF_v|_{\widetilde{U}_1})$. If
$s : V\to [L]$ is $\PP$-constant, $s$ can be decomposed as
$s(a,b)=f(\alpha)\cdot s_0(a,b)$, where $f$ is a function on
$V/\PP=\SS^2$ and $\alpha$ is the midpoint of the short geodesic
connecting $a$ to $b$. But then, $s$ can equally well be
decomposed as $s(a,b)=\widetilde{f}(-\alpha)\cdot s_1(a,b)$, where
$\widetilde{f}$ is also a function on $V/\PP=\SS^2$, since
$-\alpha$ is the midpoint of the long geodesic connecting $a$ to
$b$. It follows from the above that we must have:
$$
\widetilde{f}(-\alpha)=(-1)^n\cdot f(\alpha) \ , \
\forall\alpha\in\SS^2 \ . \tag{6}
$$

Therefore, to any $\PP$-constant section $s : V\to [L]$, we
associate two functions $f$ and $\widetilde{f}$ on $V/\PP=\SS^2$
satisfying \thetag{6}. Clearly, \thetag{6} is a very strong
relation, for, if $g$ is a function of definite parity with
respect to the antipodal mapping, i.e., if $g(-\alpha)=(-1)^k\cdot
g(\alpha)$, $k$ integer, then, from \thetag{6} it follows that
$\widetilde{g}(-\alpha)=(-1)^n\cdot g(\alpha)=(-1)^n\cdot
(-1)^k\cdot g(-\alpha)=(-1)^k\cdot \widetilde{g}(\alpha)$, so that
$\widetilde{g}$ has the same parity as $g$ and furthermore
$\widetilde{g}=\pm g$ is such that
$$
\widetilde{g}=g \ \Leftrightarrow \ g(\alpha)=(-1)^n\cdot
g(-\alpha) \ , \tag{7a}
$$
$$\widetilde{g}=-g \ \Leftrightarrow \ g(\alpha)=-(-1)^n\cdot g(-\alpha) \ .
\tag{7b}
$$

Accordingly, the space of functions on $\SS^2$ satisfying
\thetag{7a} or \thetag{7b}, respectively, will be called the {\it
$n$-even} or {\it $n$-odd} subspace of ${\Cal Fun}(\SS^2)$ and
denoted by ${\Cal Fun}^{n}_{+}(\SS^2)$ or ${\Cal
Fun}^{n}_{-}(\SS^2)$, respectively. For every $n\in {\bold
Z^{+}}$, we have ${\Cal Fun}(\SS^2)$ = ${\Cal Fun}^{n}_{+}(\SS^2)
\oplus {\Cal Fun}^{n}_{-}(\SS^2)$.

Thus, if we concentrate on functions of definite parity with
respect to the antipodal map, then it is clear from \thetag{7}
that in practice we associate a single function $g$ to any
$\PP$-constant section $s : V\to [L]$ of definite parity. Since
functions without definite parity can always be decomposed into
such, if $g_{+}$ or $g_{-}$ satisfies \thetag{7a} or \thetag{7b},
respectively, then
$$
f=g_{+}+g_{-} \ \Leftrightarrow \ \widetilde{f}=g_{+}-g_{-} \ .
\tag{8}
$$
In this way, we uniquely associate a single pair of $n$-even and
$n$-odd functions $(g_+,g_-)$ on $\SS^2$ to any $\PP$-constant
section $s : V\to [L]$, with the two choices of associating this
pair of functions to a polarized section, as presented above.

Continuing the construction, we  now generalize the {\it
composition of central polarized sections}. Remember the sequence
of steps: we start with two polarized sections $s^1$ and $s^2$,
multiply them to get a new unpolarized section $s^1 \circledcirc
s^2$, then average over the fibers to get a final central
polarized, or $\PP$-constant section $<s^1 \circledcirc s^2>$.

When the association of polarized sections with functions is
unique, as in spaces without closed geodesics, this procedure
yields a unique skewed product of functions. However, now to any
polarized section $s^1$ we have two choices of association:
$s^1(a,b) = f_1(\alpha)\cdot s_0(a,b)$ and $s^1(a,b) =
\widetilde{f}_1(-\alpha)\cdot s_1(a,b)$, with $f$ and
$\widetilde{f}$ related by \thetag{6} and \thetag{8}.

Thus, the skewed product of functions associated to $<s^1
\circledcirc s^2>$ can now be written in $2^3=8$ ways. But it is
not difficult to see that \thetag{6} and \thetag{8} guarantee that
these are all equivalent, in accordance with the uniqueness of
$<s^1 \circledcirc s^2>$.

\subhead Eight partial skewed products of functions on the sphere
\endsubhead

We have just seen that if one starts with two central polarized
sections $s^1 , s^2 : V\to [L]$, their composition is uniquely
defined. However, if one starts with two functions $f_1 , f_2$ on
the sphere, there would seem to be a greater freedom in obtaining
their skewed product.

More specifically, given a generic function $f_1$ on $\SS^2$,  we
can generically associate two distinct polarized sections
$[s^1]_0$ and $[s^1]_1$ from $V$ to $[L]$ by $[s^1]_0(a,b) =
f_1(\alpha)\cdot s_0(a,b)$ and $[s^1]_1(a',b') = f_1(\alpha)\cdot
s_1(a',b')$. In this context, the two ``types'' of sections, $0$
or $1$, actually refer to two distinct central polarized sections.
No equivalence condition, at this point.

Therefore, given two generic functions $f_1$ and $f_2$ on $\SS^2$,
there are now $4$ distinct central polarized sections to be
multiplied. Each of these product sections can be of type $0$ or
$1$ and, since we are now taking the functions themselves to be
the basic entities, each type of product section, for each of the
$4$ products, could define a distinct function on $\SS^2$.

It follows that there are, in principle, $8$ different ways of
obtaining a partial skewed product $f_1 \star f_2$ and that these
$8$ products of functions are not necessarily the same. Instead,
these $8$ products divide into $4$ groups of ``conjugate'' pairs
of products, which are:
$$
\aligned &(f_1 \star f_2)_{000} \ \& \ (f_1 \star f_2)_{111} \ ,
\\
&(f_1 \star f_2)_{001} \ \& \ (f_1 \star f_2)_{110} \ ,
\\
&(f_1 \star f_2)_{010} \ \& \ (f_1 \star f_2)_{101} \ ,
\\
&(f_1 \star f_2)_{100} \ \& \ (f_1 \star f_2)_{011} \ ,
\endaligned
\tag{9}
$$
according to the $8$ ways of composing polarized sections
associated to $f_1$ and $f_2$:
$$
\aligned &<[[s^1]_0\circledcirc [s^2]_0]_0> \ \& \
<[[s^1]_1\circledcirc [s^2]_1]_1> \ ,
\\
&<[[s^1]_0\circledcirc [s^2]_0]_1> \ \& \ <[[s^1]_1\circledcirc
[s^2]_1]_0> \ ,
\\
&<[[s^1]_0\circledcirc [s^2]_1]_0> \ \& \ <[[s^1]_1\circledcirc
[s^2]_0]_1> \ ,
\\
&<[[s^1]_1\circledcirc [s^2]_0]_0> \ \& \ <[[s^1]_0\circledcirc
[s^2]_1]_1> \ .
\endaligned
\tag{10}
$$

Here, $<[[s^1]_1\circledcirc [s^2]_0]_1>$ stands for the following
operation: $f_1$ is associated to a polarized section $[s^1]_1$
while $f_2$ is associated to $[s^2]_0$ ; these polarized sections
multiply (see \cite{RO}) into an unpolarized section of ``type''
$1$ that is $[[s^1]_1\circledcirc [s^2]_0]_1$ , which after
averaging over the fibers becomes a polarized section
$<[[s^1]_1\circledcirc [s^2]_0]_1> $ , to which is associated the
function $(f_1 \star f_2)_{101}$ . Similarly for all other cases.
The ``standard'' case $<[[s^1]_0\circledcirc [s^2]_0]_0>$ with
$(f_1 \star f_2)_{000}$ is the case that was considered
previously, resulting in equation \thetag{1}.

Thus, the construction of each ``nonstandard'' partial product
$(f_1 \star f_2)_{\eta\nu\rho}$ , where $\eta,\nu,\rho,  \in \{
0,1 \}$, is a natural generalization of the standard one so that
it can be written as:
$$
(f_1 \star f_2)_{\eta\nu\rho}(m) = \iint_{W_{m}^{\eta\nu\rho}}
f_1(m')f_2(m'')K_{\eta\nu\rho}(m',m'',m)dm'dm'' \ ,
$$
where each integral kernel $K_{\eta\nu\rho}(m',m'',m) =
A_{\eta\nu\rho}(m',m'',m)\e^{in\Sn_{\eta\nu\rho}(m',m'',m)/2}$.

From the definition of the polarized sections $s_0$ and $s_1$ it
follows that $\Sn_{\eta\nu\rho}(m',m'',m)$ is the symplectic area
of the geodesic triangle with midpoints $m',m'',m$, whose side
through $m'$ is short ($<\pi$) if $\eta=0$ or long ($>\pi$) if
$\eta=1$, and so on for the others.

Similarly, $W_{m}^{\eta\nu\rho}\subset\SS^2\times\SS^2$ is the
third restriction (to $m$) of
$W^{\eta\nu\rho}\subset\SS^2\times\SS^2\times\SS^2$ and this can
be identified as the space of geodesic triangles as determined by
the midpoints, whose side through the first midpoint is short
($<\pi$) if $\eta=0$ or long ($>\pi$) if $\eta=1$, and so on for
the others. Accordingly, $A_{\eta\nu\rho}(m',m'',m)dm'dm''dm$ is
the natural volume form on this space, obtained similarly to the
standard case.

However, aside from a set of measure zero in $W^{\eta\nu\rho}$,
each nonstandard geodesic triangle
$\overline{\Delta}^{\eta\nu\rho}(m',m'',m) \in W^{\eta\nu\rho}$
has well defined vertices $a,b,c\in \SS^2$ and corresponds
uniquely to a standard geodesic triangle
$\overline{\Delta}^{000}(\widetilde{m'},\widetilde{m''},\widetilde{m})$,
where $\widetilde{m'}=m'$ if $\eta=0$, or $\widetilde{m'}=-m'$ if
$\eta=1$, and so on for the others. It follows that each
$W^{\eta\nu\rho}$ is isomorphic to $W^{000}\equiv W$ and that
their volume forms are the same, that is, $A_{\eta\nu\rho}\equiv
A$ is given by formula \thetag{4}, because each change
$m'\to\widetilde{m'}$, etc, could only multiply $A$ by $\pm 1$,
but $A_{\eta\nu\rho}$ is a nonnegative function.

\subhead The global skewed product of functions on the sphere
\endsubhead

So, what is to be done of these $8$ partial skewed products of
functions? At this point, it is important to remember how each
partial skewed product is defined:

Starting with two functions on the sphere, each function is
associated to a central polarized section and these sections are
multiplied into a new section that is {\it not} polarized. This
unpolarized section is associated, not to a function on the
sphere, but to a continuous family of functions on the sphere,
parametrized by the points on each leaf of $\PP$, or in other
words, each vector in (a subset of) the tangent space over a point
on the sphere.

It is only after {\it averaging} over this continuous family of
functions on the sphere, or in other words, over each leaf of
$\PP$, or similarly over each (subset of the) tangent space over a
point, that a new function on the sphere is defined. In this way,
the {\it averaging procedure} is fundamental to the construction
of each partial skewed product of functions.

Now, we have just found out that, starting with two generic
function on the sphere, we have further obtained a discrete family
of functions on the sphere, according to the eight possible ways
of producing a partial skewed product. Therefore, in order to
obtain a single function on the sphere, the natural thing to do is
to {\it average} over this family of functions, naturally
generalizing the {\it averaging procedure} to this discrete
family.

In order to do this, we start by focusing attention momentarily on
the first column of \thetag{9} and \thetag{10}. There, we note
that all geodesic triangles involved have at most one side that is
long. For such triangles, formula \thetag{2} applies directly
\cite{RO}, so that
$$ \Sn_{\eta\nu\rho}\equiv
\Sn \ (formula \ \thetag{2}) \ , \ \forall(\eta\nu\rho)\in\{
(000),(001),(010),(100)\} \ . \tag{11}
$$

Furthermore, if $(\eta\nu\rho)\in\{ (000),(001),(010),(100)\}$, by
direct inspection we also note that the four $W_m^{\eta\nu\rho}$
are mutually disjoint, except for sets of measure zero, and
$W_m^{000}\cup W_m^{001}\cup W_m^{010}\cup W_m^{100} =
\SS^2\times\SS^2$. This is seen from the spherical trigonometric
relation
$\cos(y_1)/\cos(z_1)=\cos(y_2)/\cos(z_2)=\cos(y_3)/\cos(z_3)$,
where $y_1$ is half the length of side $1$ and $z_1$ is the
distance between the midpoints of the other sides, and so on
\cite{RO}. Thus, $W^{001}$ is determined by: $\sign\euc {m'}{m} =
\sign \euc {m''}m = -\sign \euc {m'}{m''}$, and so on.

It follows from all of the above that we can average the four
partial products obtained in the left column of \thetag{9},
associated to the left column of \thetag{10}, into a single global
product defined on $\SS^2\times\SS^2$ which we shall denote as
$(f_1 \star f_2)_{[\bold{0}]}$ and is given by:
$$
(f_1 \star f_2)_{[\bold{0}]}(m) = \frac{1}{4}
\iint_{\SS^2\times\SS^2}
f_1(m')f_2(m'')A(m',m'',m)\e^{in\Sn(m',m'',m)/2}dm'dm'' \ ,
$$
where $\Sn$ and $A$ are given by formulas \thetag{2} and
\thetag{4}, respectively.

This is still only half of the story, of course, because we also
have to take into account the right column of \thetag{9} and
\thetag{10}. To do so, we must understand precisely the relation
between {\it conjugate} midpoint triangles, that is, geodesic
triangles with the same midpoints (which do not belong to a
degenerate set).

More explicitly, if $(m',m'',m)\in W^{000}$ determines a unique
standard geodesic triangle with vertices $(a,b,c)$, then
$(m',m'',m)$ also determine a geodesic triangle with three long
sides and vertices $(-a,-b,-c)$. To see this, just prolong the
short geodesic $\overline{ab}$ to a long geodesic
$\overline{-a-b}$, and so on. The short triangle with vertices
$(a,b,c)$ and the long triangle with vertices $(-a,-b,-c)$ have
the same midpoints $(m',m'',m)$ so they are conjugate to each
other. It follows immediately that $W^{000}\equiv W^{111}$.
Similarly, $W^{001}\equiv W^{110}$, $W^{010}\equiv W^{101}$ and
$W^{100}\equiv W^{011}$. Thus each pair of conjugate products in
each line of \thetag{9} is defined in a same domain
$W_m^{\eta\nu\rho}\subset\SS^2\times\SS^2$, with the same
amplitude function $A$.

As for the holonomy of each composition, since it does not depend
on $f_1$ or $f_2$ and since $s_1(-a,-b)=(-1)^n\cdot s_0(-a,-b)$,
then, starting with $(m',m'',m)\in W^{000}$, clearly the holonomy
over the long triangle with vertices $(-a,-b,-c)$ is equal to
$(-1)^n$ times the holonomy over the short triangle with these
same vertices, whose midpoints are $(-m',-m'',-m)$, which is
written as $e^{in\Sn(-m',-m'',-m)/2}$, where $\Sn$ is given by
equation \thetag{2}. But from \thetag{2}, $\eta(-m',-m'',-m) =
\eta(m',m'',m)$ and $det(-m',-m'',-m) = -det(m',m'',m)$, so
$$
e^{in\Sn(-m',-m'',-m)/2} = e^{-in\Sn(m',m'',m)/2} \ . \tag{12}
$$
Therefore, the holonomy over the long triangle whose midpoints are
$(m',m'',m)$ is obtained from the holonomy over the short triangle
with the same midpoints by
$$
e^{in\Sn_{111}(m',m'',m)/2} = (-1)^n\cdot
e^{-in\Sn_{000}(m',m'',m)/2} \ .
$$
And the same analysis applies when comparing any pair of conjugate
products, i.e., when comparing the right to the left column in
each line of \thetag{9} or \thetag{10}, that is:
$$
\aligned e^{in\Sn_{\bar{\eta}\bar{\nu}\bar{\rho}}(m',m'',m)/2} =
(-1)^n\cdot e^{-in\Sn_{\eta\nu\rho}(m',m'',m)/2} \ ,
\\
\eta+\bar{\eta}=\nu+\bar{\nu}=\rho+\bar{\rho}=1 \ (modulo \ 2) \ .
\endaligned
\tag{13}
$$

It follows from the above that we can average the $4$ products
obtained in the right column of \thetag{9}, associated to the
right column of \thetag{10}, into another single global product
defined on $\SS^2\times\SS^2$ which we shall denote as $(f_1 \star
f_2)_{[\bold{1}]}$ and is given by:
$$
(f_1 \star f_2)_{[\bold{1}]}(m) = \frac{(-1)^n}{4}
\iint_{\SS^2\times\SS^2}
f_1(m')f_2(m'')A(m',m'',m)\e^{-in\Sn(m',m'',m)/2}dm'dm'' \ ,
$$
where again $A$ and $\Sn$ are given by formulas \thetag{4} and
\thetag{2}, respectively.

Note that $(f_1 \star f_2)_{[\bold{0}]}$ and $(f_1 \star
f_2)_{[\bold{1}]}$ are not the same, instead they are related by
$$
(f_1 \star f_2)_{[\bold{1}]} = (-1)^n\cdot (f_2 \star
f_1)_{[\bold{0}]} \ . \tag{14}
$$

The final step of the construction is, of course, to average the
two global products which were obtained by averaging each of the
columns in \thetag{9}. In this way, we arrive at {\it the global
skewed product} of functions on the sphere, denoted by $\star^n$,
which is the average over all partial skewed products on the
sphere, given by ($\eta,\nu,\rho\in\{0,1\}$):
$$
 f_1\star^n f_2   = \frac{1}{8} \sum_{\eta\nu\rho} (f_1 \star
f_2)_{\eta\nu\rho}  = \frac{1}{2}  \left\{ \ (f_1 \star
f_2)_{[\bold{0}]} + (f_1 \star f_2)_{[\bold{1}]} \ \right\} \ .
\tag{15}
$$

We say that the global skewed product is ${\bold Z_2}$-{\it graded
commutative} because
$$
f_1\star^n f_2 = (-1)^n \ f_2\star^n f_1 \ , \tag{16}
$$
as follows immediately from \thetag{14} and \thetag{15}. Our
results are summarized below:

\

\noindent{\bf Theorem 1}: {\it For the $2$-sphere, the partial
skewed product of functions given by formula \thetag{1}, where
$n=2/\hbar \in {\bold Z^{+}}$, with $A$ and $S$ given by formulas
\thetag{4} and \thetag{2}, respectively, extends naturally to the
global skewed product $\star^n$ which is uniquely obtained by
averaging over the eight possible partial skewed products of
functions on the sphere. The global skewed product is ${\bold
Z_2}$-graded commutative \thetag{16} and is given explicitly by
($k\in {\bold Z^{+}}$)}:
$$
f_1\star^{2k} f_2 \ (m) = \iint_{\SS^2\times\SS^2}
f_1(m')f_2(m'')\frac{1}{4}A(m',m'',m)\cos\{k\Sn(m',m'',m)\}
dm'dm'' \ , \tag{17a}
$$
$$
f_1\star^{2k-1} f_2 \ (m) = \iint_{\SS^2\times\SS^2}
f_1(m')f_2(m'')\frac{i}{4}A(m',m'',m)\sin\{
(k-\frac{1}{2})\Sn(m',m'',m) \} dm'dm'' \tag{17b}
$$

\

\subhead Skewed products of functions with definite parity
\endsubhead

We now focus attention on functions with definite parity, i.e.
$n$-even functions in ${\Cal Fun}^{n}_{+}(\SS^2)$ and $n$-odd
functions in ${\Cal Fun}^{n}_{-}(\SS^2)$ , satisfying \thetag{7a}
and \thetag{7b}, respectively.

Start with a $n$-even function $g_1 \in {\Cal
Fun}^{n}_{+}(\SS^2)$. For such a function, $\widetilde{g_1}=g_1$
so that the two polarized sections $[s^1]_0$ and $[s^1]_1$ from
$V$ to $[L]$ given by $[s^1]_0(a,b) = g_1(\alpha)\cdot s_0(a,b)$
and $[s^1]_1(a',b') = g_1(\alpha)\cdot s_1(a',b')$ are actually
the same. Therefore, if we multiply two such functions, it is
particularly important to understand how all partial products
relate to each other and to the global product given by formula
\thetag{17}.

It turns out that, in order to compare the partial products, it is
necessary to express how the midpoint triangular area
$S(m',m'',m)$ changes as some of its arguments are changed under
the antipodal mapping. From formula \thetag{2}, if only one of the
arguments is changed (say $(m',m'',m)\mapsto (m',m'',-m)$, for
instance), then it is clear that $\eta\mapsto -\eta$ and $det
\mapsto -det$, so that $S/2\mapsto S/2 \pm \pi$ and therefore
$$
e^{in\Sn(m',m'',-m)/2} = (-1)^n\cdot e^{in\Sn(m',m'',m)/2} \ .
\tag{18}
$$
On the other hand, if two of the arguments are changed (for
instance, $(m',m'',m)\mapsto (-m',-m'',m)$), then $\eta\mapsto
-\eta$ but $det \mapsto det$, so that
$$
e^{in\Sn(-m',-m'',m)/2} = (-1)^n\cdot e^{-in\Sn(m',m'',m)/2} \ .
\tag{19}
$$
Finally, combining these two transformations we recover the case
for when the three arguments are changed, as previously discussed,
which is given by formula \thetag{12}.

Now, let's start by comparing $(g_1 \star g_2)_{010}$ to $(g_1
\star g_2)_{000}$, when $g_1,g_2\in {\Cal Fun}^{n}_{+}(\SS^2)$.
From
$$
(g_1 \star g_2)_{010}(m) = \iint_{W_{m}^{010}}
g_1(m')g_2(m'')A(m',m'',m)\e^{in\Sn(m',m'',m)/2}dm'dm'' \ ,
$$
we notice that, if $(m',m'')\in W_{m}^{010}$ then $(m',-m'')\in
W_{m}^{000}$, so that $\iint_{W_{m}^{010}}dm'dm''... =
\iint_{W_{m}^{000}} dm'd(-m'')...$ and thus, using formulas
\thetag{7a} and \thetag{18}, we can write
$$
(g_1 \star g_2)_{010}(m) = \iint_{W_{m}^{000}}
g_1(m')g_2(-m'')A(m',-m'',m)\e^{in\Sn(m',-m'',m)/2}dm'd(-m'') \ ,
$$
so that $(g_1 \star g_2)_{010}\equiv (g_1 \star g_2)_{000}$. A
similar analysis shows that $(g_1 \star g_2)_{100}\equiv (g_1
\star g_2)_{000}$.

We now compare $(g_1 \star g_2)_{001}$ to $(g_1 \star g_2)_{000}$.
Starting from
$$
(g_1 \star g_2)_{001}(m) = \iint_{W_{m}^{001}}
g_1(m')g_2(m'')A(m',m'',m)\e^{in\Sn(m',m'',m)/2}dm'dm'' \ ,
$$
we notice that, if $(m',m'')\in W_{m}^{001}$ then $(-m',-m'')\in
W_{m}^{000}$, so that $\iint_{W_{m}^{001}}dm'dm''... =
\iint_{W_{m}^{000}} d(-m')d(-m'')...$ and thus, using formulas
\thetag{7a} and \thetag{19}, we write $(g_1 \star g_2)_{001}(m) =$
$$
= (-1)^n\cdot\iint_{W_{m}^{000}}
g_1(-m')g_2(-m'')A(-m',-m'',m)\e^{-in\Sn(-m',-m'',m)/2}d(-m')d(-m'')
\ ,
$$
so that $(g_1 \star g_2)_{001}\equiv (-1)^n\cdot (g_2 \star
g_1)_{000}$.

On the other hand, if $(m',m'',m)\in W^{001}$ then $(m',m'',-m)\in
W^{000}$, so that $W_{m}^{001}\equiv W_{-m}^{000}$, hence
$\iint_{W_{m}^{001}}dm'dm''... = \iint_{W_{-m}^{000}} dm'dm''...$
and thus, using formula \thetag{18}, we write
$$
(g_1 \star g_2)_{001}(m) = (-1)^n\cdot\iint_{W_{-m}^{000}}
g_1(m')g_2(m'')A(m',m'',-m)\e^{in\Sn(m',m'',-m)/2}dm'dm'' \ ,
$$
yielding $(g_1 \star g_2)_{001}(m) = (-1)^n\cdot (g_1 \star
g_2)_{000}(-m)$. Combining the two results we have that
$$
(g_2 \star g_1)_{000}(m) = (g_1 \star g_2)_{000}(-m) \ ,
\tag{20}
$$
a formula that can also be obtained directly from formulas
\thetag{1}, \thetag{7a} and \thetag{12}.

Now, if we follow through with the averaging procedure, we obtain
from formula \thetag{13} that $(g_1 \star g_2)_{111}\equiv
(-1)^n\cdot (g_2 \star g_1)_{000}$ and, combining with the
previous results, we obtain from formula \thetag{13} that $(g_1
\star g_2)_{101}\equiv (-1)^n\cdot (g_2 \star g_1)_{010}\equiv
(-1)^n\cdot (g_2 \star g_1)_{000}$, also $(g_1 \star
g_2)_{011}\equiv (-1)^n\cdot (g_2 \star g_1)_{100}\equiv
(-1)^n\cdot (g_2 \star g_1)_{000}$ and finally $(g_1 \star
g_2)_{110}\equiv (-1)^n\cdot (g_2 \star g_1)_{001}\equiv (g_1
\star g_2)_{000}$. Therefore, averaging all partial products
yields:
$$
g_1\star^n g_2  = \frac{1}{8} \sum_{\eta\nu\rho} (g_1 \star
g_2)_{\eta\nu\rho} = \frac{1}{2} \left\{ (g_1 \star g_2)_{000} +
(-1)^n\cdot (g_2 \star g_1)_{000} \right\} \tag{21}
$$
and it follows immediately from \thetag{20} (and \thetag{7a}) that
$g_1\star^n g_2  \in {\Cal Fun}^{n}_{+}(\SS^2)$. Explicitly:
$$
g_1\star^n g_2 \ (m) = \frac{1}{2} \left\{ (g_1 \star
g_2)_{000}(m) + (-1)^n\cdot (g_1 \star g_2)_{000}(-m) \right\} \ ,
\tag{22}
$$
or even more explicitly, with $W_m \equiv W_m^{000}$ given by
formula \thetag{3} and for $k \in {\bold Z^{+}}$:
$$
g_1\star^{2k} g_2 \ (m) = \iint_{W_m}
g_1(m')g_2(m'')A(m',m'',m)\cos\{k\Sn(m',m'',m)\} dm'dm'' \ ,
\tag{23a}
$$
$$
g_1\star^{2k-1} g_2 \ (m) = \iint_{W_m}
g_1(m')g_2(m'')iA(m',m'',m)\sin\{ (k-\frac{1}{2})\Sn(m',m'',m) \}
dm'dm'' \tag{23b}
$$

Of course, it also follows directly from formulas \thetag{17} and
\thetag{18} that for any $f_1, f_2 \in {\Cal Fun}(\SS^2)$ their
global skewed product belongs to  ${\Cal Fun}^{n}_{+}(\SS^2)$.
Therefore, we now study the products of functions of definite
parity when at least one belongs to ${\Cal Fun}^{n}_{-}(\SS^2)$.

Start with $g_1, g_2 \in {\Cal Fun}^{n}_{-}(\SS^2)$. Similar
considerations, using formulas \thetag{7b}, \thetag{12},
\thetag{18} and \thetag{19}, show that in this case $(g_1 \star
g_2)_{010}\equiv -(g_1 \star g_2)_{000}$ and $(g_1 \star
g_2)_{100}\equiv -(g_1 \star g_2)_{000}$, while again $(g_1 \star
g_2)_{001}\equiv (-1)^n\cdot (g_2 \star g_1)_{000}$. From formula
\thetag{13} we again obtain that $(g_1 \star g_2)_{111}\equiv
(-1)^n\cdot (g_2 \star g_1)_{000}$, while $(g_1 \star
g_2)_{101}\equiv (-1)^n\cdot (g_2 \star g_1)_{010}\equiv
-(-1)^n\cdot (g_2 \star g_1)_{000}$, also $(g_1 \star
g_2)_{011}\equiv (-1)^n\cdot (g_2 \star g_1)_{100}\equiv
-(-1)^n\cdot (g_2 \star g_1)_{000}$ and finally $(g_1 \star
g_2)_{110}\equiv (-1)^n\cdot (g_2 \star g_1)_{001}\equiv (g_1
\star g_2)_{000}$. It follows that these eight products average to
zero, that is, if $g_1, g_2 \in {\Cal Fun}^{n}_{-}(\SS^2)$ then $
g_1\star^n g_2  \equiv 0$.

If $g_1 \in {\Cal Fun}^{n}_{+}(\SS^2)$ and $g_2 \in {\Cal
Fun}^{n}_{-}(\SS^2)$, then we obtain that $(g_1 \star
g_2)_{010}\equiv -(g_1 \star g_2)_{000}$, but $(g_1 \star
g_2)_{100}\equiv (g_1 \star g_2)_{000}$ and $(g_1 \star
g_2)_{001}\equiv -(-1)^n\cdot (g_2 \star g_1)_{000}$. Also, $(g_1
\star g_2)_{111}\equiv (-1)^n\cdot (g_2 \star g_1)_{000}$ and
$(g_1 \star g_2)_{101}\equiv (-1)^n\cdot (g_2 \star
g_1)_{010}\equiv -(-1)^n\cdot (g_2 \star g_1)_{000}$, but $(g_1
\star g_2)_{011}\equiv (-1)^n\cdot (g_2 \star g_1)_{100}\equiv
(-1)^n\cdot (g_2 \star g_1)_{000}$ and finally $(g_1 \star
g_2)_{110}\equiv (-1)^n\cdot (g_2 \star g_1)_{001}\equiv -(g_1
\star g_2)_{000}$. Thus, again, these eight products average to
zero. Similarly if $g_1 \in {\Cal Fun}^{n}_{-}(\SS^2)$ and $g_2
\in {\Cal Fun}^{n}_{+}(\SS^2)$. In both cases, $g_1\star^n g_2
\equiv 0$. In summary:

\

\noindent{\bf Theorem 2}: {\it The global skewed product on ${\Cal
Fun}(\SS^2)$, which is given by formula \thetag{17}, is nontrivial
only as a product of $n$-even functions: ${\Cal
Fun}^{n}_{+}(\SS^2) \times {\Cal Fun}^{n}_{+}(\SS^2) \to {\Cal
Fun}^{n}_{+}(\SS^2)$, so that ${\Cal Fun}(\SS^2) \times {\Cal
Fun}^{n}_{-}(\SS^2) \to 0$ and ${\Cal Fun}^{n}_{-}(\SS^2) \times
{\Cal Fun}(\SS^2) \to 0$. Furthermore, on ${\Cal
Fun}^{n}_{+}(\SS^2)\subset{\Cal Fun}(\SS^2)$, the subspace of
functions satisfying \thetag{7a}, the global skewed product
coincides with the restricted skewed product given by formulas
\thetag{21}, \thetag{22} and \thetag{23}, which is the ${\bold
Z_2}$-graded commutative version of the partial skewed product
given by formula \thetag{1}}.

\subhead Generalized skewed products of functions on the sphere
\endsubhead

We have seen that although the composition of central polarized
sections is uniquely defined, the corresponding skewed product of
functions seemed to be non unique. This apparent lack of
uniqueness of the skewed product would originate in the lack of
uniqueness in associating, to each function, a central polarized
section. Thus, in order to obtain a unique skewed product of
functions we averaged over all possibilities, in line with the
averaging procedure that was used to define each partial skewed
product. In so doing, we have obtained a global skewed product
which is ${\bold Z_2}$-graded commutative.

However, for functions in ${\Cal Fun}^{n}_{+}(\SS^2)$, that is,
functions of definite parity satisfying formula \thetag{7a}, there
is no lack of uniqueness in associating, to each function, a
central polarized section. Therefore, it would be only natural
that their skewed product should be uniquely associated to a
central polarized section, in other words, that their skewed
product should also belong to ${\Cal Fun}^{n}_{+}(\SS^2)$. This is
guaranteed by averaging, for the composition of central polarized
sections, the two ways of associating a function to the composed
section, according to formula \thetag{6}. This is exactly what is
explicitly stated by formula \thetag{22}, which, furthermore,
asserts the equivalence between all other pairs of skewed products
and, specially, the equivalence of any well-defined skewed product
on ${\Cal Fun}^{n}_{+}(\SS^2)$ to the global skewed product of
functions, which, for all practical purposes, is also defined only
on ${\Cal Fun}^{n}_{+}(\SS^2)$.

In this way, we have seen that the two ways of assuring that the
skewed product of functions on the sphere is unique and well
defined, turn out to be the same.

Furthermore, we note that the specific form of the amplitude
function (formula \thetag{4}) is not relevant to the ${\bold
Z_2}$-graded commutativity of the skewed product of functions on
the sphere. As long as the amplitude function is a real
nonnegative symmetric function (formula \thetag{5}) which is
invariant under the $SU(2)$ diagonal action on
$\SS^2\times\SS^2\times\SS^2$, it is the relation between
holonomies over conjugate midpoint triangles (formula \thetag{13})
that implies the skewed product to be nontrivial only on ${\Cal
Fun}^{n}_{+}(\SS^2)$ and to be ${\bold Z_2}$-graded commutative.

Now, if another integral product of functions on $\SS^2$ is to be
defined so that the amplitude function is different from the one
defining the skewed product (and could depend on $n$, but not
exponentially, say, polinomially in $1/n$), but such that the
phase of the oscillatory function is defined via midpoint
triangles on $\SS^2$, then, since for any triple of points in
$\SS^2$ there is no a priori reason to choose one midpoint
triangle over another and so all midpoint triangles should be
considered equally, this implies the more general result stated
below:

\

\noindent{\bf Corollary 3}: {\it Given $n \in {\bold Z^{+}}$, the
skewed product $\star^n$ on ${\Cal Fun}(\SS^2)$ is well defined
only because it is ${\bold Z_2}$-graded commutative, $f\star^n g =
(-1)^ng\star^n f$, and is nontrivial only as a product on the
subspace ${\Cal Fun}^{n}_{+}(\SS^2)$ of functions satisfying
$f(m)=(-1)^nf(-m)$, where $-m$ is antipodal to $m$. This is also
true for any generalized skewed product $\tilde{\star}^n$ of the
form $f \ \tilde{\star}^n g \ (m) = \iint_{\SS^2\times\SS^2}
\tilde{K}_n(m,m',m'')f(m')g(m'')dm'dm''$, where the integral
kernel $\tilde{K}_n$ is symmetric (formula \thetag{5}) and $SU(2)$
invariant, of the form $\tilde{K}_n = \tilde{A}_n\exp{(in{\Cal
S}/2)}$, where $\tilde{A}_n$ is a nonnegative real function on
$\SS^2\times\SS^2\times\SS^2$ (possibly expanded in powers of
$1/n$) and ${\Cal S}(m,m',m'')$ is the symplectic area of a
geodesic triangle with $(m,m',m'')$ as midpoints. Because
$\exp{(in{\Cal S}/2)}$ is double valued (formula \thetag{13}) and
counting all possibilities equally, then $\tilde{K}_{2k} =
\tilde{A}_{2k}\cos{(k\Sn)}$ and $\tilde{K}_{2k-1} =
i\tilde{A}_{2k-1}\sin{((k-\frac{1}{2})\Sn)}$, with $\Sn$ given by
formula \thetag{2}, which implies that $\tilde{\star}^n$ is
nontrivial only on ${\Cal Fun}^{n}_{+}(\SS^2)$ and is ${\bold
Z_2}$-graded commutative}.

\head Discussion \endhead

Formal star products on the sphere are not ${\bold Z_2}$-graded
commutative, neither commutative \cite{MO}\cite{Md}. However, an
old result by Rieffel \cite{Rf}, based on a theorem by Wassermann
\cite{Wa}, asserts that any associative $SU(2)$-equivariant {\it
strict} deformation of the pointwise product of functions on the
sphere must be commutative.

A $SU(2)$-equivariant formal star product on the sphere does not
satisfy Rieffel's strict deformation conditions because it is a
product of formal power series in $\hbar$ with coefficients in the
space of smooth functions on the sphere, which, however, generally
does not converge as power series and, therefore, is not defined
as a product in any infinite dimensional normed ${\Cal
C}^*$-algebra which deforms the usual ${\Cal C}^*$-algebra of
functions on the sphere.

On the other hand, for any $n=2/\hbar\in{\bold Z^{+}}$, the skewed
product is, in principle, defined as a product on some infinite
dimensional subspace of the space of {\it functions} on the
sphere, ${\Cal Fun}(\SS^2)$, which, in principle, could be
identified as the same space for when $n=\infty$ ($\hbar=0$).
However, the skewed product on ${\Cal Fun}(\SS^2)$ is ${\bold
Z_2}$-graded commutative and is nontrivial only as a product on
${\Cal Fun}^{n}_{+}(\SS^2)$. So, how does it relate to Rieffel's
theorem?

Well, although it is ${\bold Z_2}$-graded commutative, it is a
deformation ($n$ close to $\infty$) of the pointwise product of
functions on the sphere only when it is commutative ($n=2k$,
$k\in{\bold Z^{+}}$), in which case it is a deformation of the
usual ${\Cal C}^*$-subalgebra of functions on the sphere
satisfying $f(m) = f(-m)$, where $-m$ is antipodal to $m$.

In other words, the skewed product of functions on the sphere
seems to provide an explicit example of Rieffel's theorem for
strict deformation of the pointwise product of functions on
$\SS^2$, for functions satisfying $f(m) = f(-m)$. Moreover, the
skewed product of functions on $\SS^2$ seems to provide a
$SU(2)$-equivariant ``anticommutative strict deformation'' of the
Poisson bracket of functions on $\SS^2$, for functions satisfying
$f(m)=-f(-m)$, given some appropriate definition of ``strict
deformation'' of anticommutative products.

Not so fast, one should argue, because we have not studied the
convergence of the skewed product and identified an appropriate
infinite dimensional subspace of ${\Cal Fun}^{n}_{+}(\SS^2)$ where
the product converges for every (even or odd) $n$. And this must
be done, for the skewed product.

However, in this respect we have much extra freedom, because, if
we approach this problem in the context of generalized skewed
products, as in Corollary $3$, we are allowed to modify the
amplitude function in great generality so as to define a rich
infinite dimensional subspace of the space of $n$-even functions
on $\SS^2$ in which a generalized skewed product converges for
every (even or odd) $n$. That is, with so much freedom, it is
natural to assume that some generalized skewed products are,
either for every even, or for every odd $n$, analytically well
defined as products on infinite dimensional subspaces of the space
of functions on the sphere satisfying either $f(m)=f(-m)$, or
$f(m)=-f(-m)$, respectively.

But, one should still question, how about associativity? Again, if
we look at this question in the context of generalized skewed
products, then, for the commutative product, we could question
whether there are some amplitude functions $\tilde{A}_n$ for which
each respective convergent product is associative for every even
$n$. But, for the anticommutative product, associativity makes no
sense and it should translate, instead, into a question about the
Jacobi identity for some $\tilde{A}_n$, for every odd $n$. But now
these questions are not so simple, because they may be tied up
with the determination of the function spaces. And, of course,
these questions are not simpler in the context of {\it the} skewed
product, properly.

However, we remind that associativity, Jacobi identity, or  other
algebraic properties play no direct role in the ${\bold
Z_2}$-graded commutativity of the skewed product, or generalized
skewed products. Therefore, the ${\bold Z_2}$-graded commutativity
of generalized skewed products proposes that Rieffel's obstruction
to $SU(2)$-equivariant strict deformation quantization is more
general in scope. In other words, it states that more general
$SU(2)$-equivariant products that deform the pointwise product in
an infinite dimensional space of functions on the sphere are
commutative. Likewise, that more general $SU(2)$-equivariant
products that deform the Poisson bracket in an infinite
dimensional space of functions on the sphere are anticommutative.
In this respect, ${\bold Z_2}$-graded commutativity being
independent of associativity or Jacobi identity, the investigation
of these or other related algebraic properties of the skewed
product, or generalized skewed products, can be carried out
independently.

On the other hand, how relevant is the special form of integral
kernel for the skewed product, or a generalized skewed product as
in Corollary $3$? Well, we argue that this form of integral
product is very relevant in the limit $n\to\infty$.

First, if we picture this limit as the limit of spheres with
smaller and smaller curvatures (for an identification of
riemannian curvature as proportional to $1/n$), we should expect
that an integral product of functions on the sphere, with
symmetric kernel, should get more and more similar to the von
Neumann - Groenewold product of functions on $\RR^2$.

Second, regardless of pictures, in this limit $n\to\infty$,
classical data should become predominant, so, for instance, one
expects that stationary phase evaluation of an integral product of
two oscillatory functions $f$ and $g$ with phases $\phi$ and
$\gamma$, respectively, should yield a new oscillatory function
$h$ whose phase $\eta$ is, at least to lowest order in $1/n$,
related to $\phi$ and $\gamma$ by the rule of composition of
central generating functions on the sphere \cite{RO}, just as
happens on $\RR^{2n}$. And so on, for multiple products one
expects the corresponding classical composition of spherical
midpoint triangles into spherical midpoint quadrilaterals, etc
\cite{RO}.

So, it is reasonable to assume that $SU(2)$-equivariant symmetric
integral products on an infinite dimensional space of functions on
the sphere should take the form of a generalized skewed product,
as $n\to\infty$. And this provides a framework for possible strict
deformation quantizations of $\SS^2$. Therefore, since we can
assume that a $SU(2)$-equivariant strict deformation quantization
of the sphere should be expressible as a generalized skewed
product in the limit $n\to\infty$ and since it is the relation
between the areas of conjugate midpoint triangles, as given by
formula \thetag{13}, that is responsible for the ${\bold
Z_2}$-graded commutativity of generalized skewed products, we
deduce that the relation between the areas of conjugate midpoint
triangles presents a simple geometrical explanation for Rieffel's
theorem on the obstruction to $SU(2)$-equivariant strict
deformation quantization of the sphere.

Furthermore, this theorem extends to a proposition on the
commutativity of more general deformations of the pointwise
product and the anticommutativity of more general deformations of
the Poisson bracket, on infinite dimensional spaces of functions
on the sphere.

On the other hand, one could feel tempted to deduce from all this
that we should forget about products on infinite dimensional
spaces of functions on the sphere and focus on formal deformation
quantization of $\SS^2$, or twisted products of spherical symbols,
only.

However, in the case $M=\RR^{2n}$, the skewed product coincides
with the twisted product and implies deformation quantization, for
admissible symbols (the Moyal product). Thus, which product (if
any) is to be given preference when $M\neq\RR^{2n}$ and the
products differ considerably? We have seen in this paper that for
$M=\SS^2$ the skewed product differs considerably from any product
in the context of formal deformation quantization, but is in line
with Rieffel's obstruction to any $SU(2)$-equivariant strict
deformation quantization of the sphere. On the other hand, both
are products motivated by mutually compatible products for
$M=\RR^{2n}$, and both are inspired by semiclassical
($n\to\infty$) considerations.

Work is in progress on the relationship between the skewed product
of functions on the sphere and twisted products on {\it finite}
dimensional spaces of spherical symbols \cite{Bn}\cite{VG}, which
are products within the context of a quantum theory, properly.

\enddocument